\documentclass[aps,pre,twocolumn,amsfonts,amssymb,amsmath,floatfix,showpacs]{revtex4}
\usepackage[dvips]{graphicx}
\usepackage{color}

\newcommand{\beps}{\varepsilon}
\newcommand{\const}{\mathrm{const}}
\newcommand{\crit}{_{\mathrm{cr}}}
\newcommand{\dt}{h_t}
\newcommand{\dx}{h_x}
\newcommand{\Estim}{E_{\mathrm{stim}}}
\newcommand{\f}{f}
\newcommand{\Heav}{\Theta}
\newcommand{\INa}{I_{\mathrm{Na}}}
\newcommand{\ms}{\mathrm{ms}}
\newcommand{\mV}{\mathrm{mV}}
\renewcommand{\O}[1]{\mathcal{O}\left(#1\right)}
\newcommand{\ustim}{u_{\mathrm{stim}}}
\newcommand{\Vstim}{V_{\mathrm{stim}}}
\newcommand{\xf}{x_f}
\newcommand{\xm}{\Delta}
\newcommand{\xstim}{x_{\mathrm{stim}}}

\newtheorem{conjecture}{Conjecture}

\def\eq(#1){(\ref{eq:#1})}
\def\eqtwo(#1,#2){(\ref{eq:#1},\ref{eq:#2})}
\def\eqthree(#1,#2,#3){(\ref{eq:#1},\ref{eq:#2},\ref{eq:#3})}
\newcommand{\Fig}[1]{Fig.~\ref{fig:#1}}
\newcommand{\fig}[1]{Fig.~\ref{fig:#1}}
\newcommand{\figs}[1]{Figs.~\ref{fig:#1}}

\begin{document}
\title{Critical fronts in initiation of excitation waves}
\author{I. Idris}
\author{V. N. Biktashev}
\affiliation{Department of Mathematical Sciences,
  University of Liverpool, Liverpool L69 7ZL, UK }
\date{\today}
\begin{abstract}
  We consider the problem of initiation of propagating wave
  in a one-dimensional excitable fiber.
  In the FitzHugh-Nagumo theory, the key role is played by
  ``critical nucleus'' and ``critical pulse'' solutions whose (center-)stable
  manifold is the threshold surface separating initial conditions
  leading to propagation and those leading to decay.
  We present evidence that in cardiac excitation models,
  this role is played by ``critical front'' solutions.
\end{abstract}
\maketitle

%##############################
\section{Introduction}

An excitable medium is a thermodynamically non-equilibrium system that
has a stable spatially uniform ``resting state'', but responds to an
above-threshold localized stimulus by a propagating non-decaying
``excitation wave''.  Excitation waves play key roles in living
organisms and are observed in chemical and physical systems,
e.g. nerves, heart muscle, catalytic redox reactions, large aspect
lasers and star formation in galaxies
\cite{Krinsky-Swinney-1991}. Understanding conditions of successful
initiation of excitation waves is particularly important for heart
where such waves trigger coordinated contraction of the muscle and
where a failure of initiation can cause or contribute to serious or
fatal medical conditions, or render inefficient the work of pacemakers
or defibrillators \cite{Zipes-Jalife-2000}.

The theoretical understanding of excitability stems from FitzHugh's
simplified model of a nerve membrane~\cite{FitzHugh-1961}. One of his
key concepts is ``quasi-threshold'', which gets precise in the limit
of large time scale separation between the processes of excitation and
recovery. Then the fast subsystem has unstable ``threshold''
equilibria. Initial conditions below such an equilibrium lead to decay,
and those above it lead to excitation.

In a spatially extended FitzHugh-Nagumo (FHN)
system~\cite{FitzHugh-1961,Nagumo-etal-1962}, the ability of a
stimulus to initiate a wave depends also on its spatial extent, the
aspect summarized by Rushton's~\cite{Rushton-1937} concept of
``liminal length''. More generic is the concept of
``critical curve'' in the stimulus strength-spatial extent plane
(see \fig{fhn}). A stimulus initiates a wave if its parameters are above this
curve, or leads to decay if below.

Mathematically, the problem is about classification of initial
conditions that will or will not lead to a traveling wave
solution. The key question is the nature of the boundary between the
two classes. A detailed analysis of this boundary has been done for
the FitzHugh-Nagumo system and its variations. This has led to the
concept of a \emph{critical nucleus}, discussed below in more
detail. Roughly, this is a spatially extended analog of a threshold
equilibrium in the point system: critical nucleus is also a stationary
but unstable solution, and its small perturbations lead to either
initiation of an excitation wave, for perturbations of one direction,
or to decay, for perturbations of the opposite direction.

We stress that although the role of FHN as a universal prototype of
excitable systems has been disputed, to our knowledge, there are still 
no alternatives to the critical nucleus concept, as far as initiation
problem is concerned.

In this paper, we present evidence that cardiac excitation provides an
example of alternative type of system, in which there is no place for
the critical nucleus. Two independent observations led to this
study. Firstly, numerical simulations of the cardiac excitation models
reveal significant qualitative differences in the way initiation
occurs in such models, compared to the FHN-style
systems~\cite{Starmer-etal-2003}. Secondly, asymptotic analysis of
detailed cardiac excitation models reveals that in the fast subsystem
there, there is \emph{no analog} of the unstable threshold
equilibrium of FHN systems~\cite{Biktashev-Suckley-2004}, and the
threshold there has a completely different mathematical nature.
Further, elementary arguments show that in cardiac equations
there are no nontrivial stationary solutions that could play the role
similar to the critical nucleus in FHN system.

Thus we have a theoretical vacuum here. Obviously, one cannot even
begin to think about investigating initiation criteria without
understanding the nature of the critical solutions. This paper aims to
fill this vacuum, and clarify the nature of the critical solutions.
We analyze a simplified model of cardiac excitation, and use the knowledge
of its exact solutions to demonstrate
that for this model the concept of critical nucleus should be replaced
with a new concept of \emph{critical front}.
We also confirm numerically the relevance of this new concept on an example of
a detailed ionic cardiac excitation model.

%##############################
\section{FitzHugh-Nagumo system}

First we recapitulate some known theoretical concepts related to
initiation of waves
(see e.g. \cite{Flores-1989,Flores-1991} and references therein).
We consider FitzHugh-Nagumo system in the form
\begin{align}
u_t &= u_{xx}+\f(u)-v , \qquad \f(u) = u(u-\theta)(1-u) \nonumber\\
v_t &= \beps (\alpha u-v)		\label{eq:fhn1}
\end{align}
where $\beps>0$, $\alpha>0$, $\theta\in(0,1/2)$
(some works consider piecewise linear functions $\f$ of similar shape)
on a half-fiber, $(x,t)\in[0,\infty)\times[0,\infty)$ with a no-flux boundary
\begin{equation}
u_x(0,t)=0 , \label{eq:fhn1b}
\end{equation}
and a rectangular
initial perturbation of width $\xstim$ and amplitude $\ustim$,
\begin{equation}
u(x,0) = \ustim\Heav(\xstim-x), \qquad
v(x,0) = 0 \label{eq:fhn1i} \\
\end{equation}
where $\Heav(\cdot)$ is the Heaviside step function.

%%%%%%%%%%%%%%%%%%
\begin{figure*}[ht]
\includegraphics{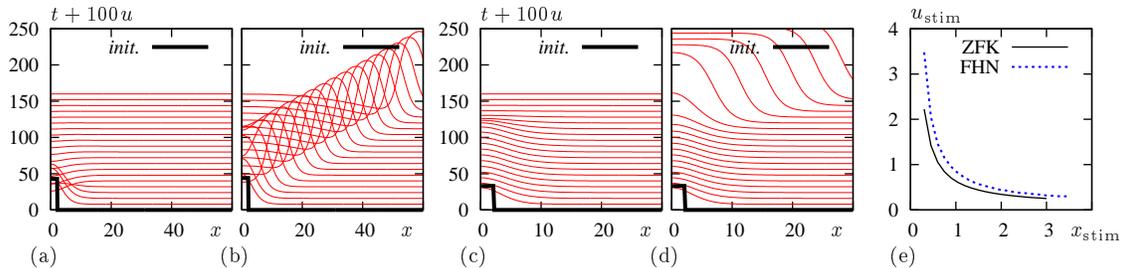}
\caption[]{
  Initiation of excitation in FitzHugh-Nagumo system.
  (a,b) Full system \eqthree(fhn1,fhn1b,fhn1i) ``FHN'' for parameter values:
  $\alpha=0.37$, $\theta=0.13$,  $\beps=0.02$.
  Stimulation parameters: $\xstim=2.10$ for both,
  below-threshold $\ustim=0.43$, leading to decay, for (a) and above-threshold $\ustim=0.44$,
  leading to initiation of excitation propagation, for (b).
  (c,d) Fast subsystem \eqthree(zfk,zfkb,zfki) ``ZFK'':
  same parameters as in (a,b) except $\beps=0$.
  Stimulation: $\xstim=2.10$ for both, below-threshold
  $\ustim=0.3304831$ for (c) and above-threshold
  $\ustim=0.3304833$ for (d).
  Bold black lines: initial conditions.
  (e) The corresponding critical curves,
  separating initiation initial conditions from decay initial conditions.
  Simulation done on an interval $x\in[0,L]$, $L=120$ with Neuman
  boundaries, central space differencing with $\dx=0.15$, and explicit
  Euler timestepping with $\dt=0.01$.
}\label{fig:fhn}
\end{figure*}
%%%%%%%%%%%%%%%%%%

\Fig{fhn}(a,b) shows two typical results of the initiation process: a
successful initiation, leading to generation of a propagating pulse,
and an unsuccessful, leading to decay of excitation in the whole
half-fiber into the resting state.

If $\beps=0$, problem
\eqthree(fhn1,fhn1b,fhn1i) reduces to an initiation problem
for Zeldovich-Frank-Kamenetsky (ZFK) equation \cite{Zeldovich-FrankKamenetsky-1938}
also known as Nagumo equation \cite{McKean-1970},
\begin{eqnarray}
u_t &=& u_{xx}+\f(u),			\label{eq:zfk}\\
u_x(0,t)&=& 0,	 			\label{eq:zfkb}\\
u(x,0) &=& \ustim\Heav(\xstim-x).	\label{eq:zfki}
\end{eqnarray}
\Fig{fhn}(c,d) illustrates the initiation and its failure in this
reduced problem. Instead of a propagating pulse, successful
initiation produces a propagating front. In the full model with small $\beps$, this front
is followed, in time scale $\O{\beps^{-1}}$, by a wave-back to
form a full excitation pulse.

A key role in understanding initiation belongs to
an unstable nontrivial bounded time-independent solution of
\eqtwo(zfk,zfkb), sometimes called \emph{critical nucleus},
by analogy with phase transitions theory.
Such solution is unique;
for a cubical nonlinearity $\f$ as in \eq(fhn1),
this solution has the form
\(
u\crit(x)=3\theta\sqrt 2%
\left[(1+\theta)\sqrt{2}+\cosh(x\sqrt \theta)\sqrt{2-5\theta+\theta^2}\right]^{-1}
\).
Its linearization spectrum has exactly one unstable
eigenvalue, while all other eigenvalues are stable. So the stable
manifold of this stationary solution has codimension one, and divides
the phase space of \eqtwo(zfk,zfkb) to two open sets. One of these
sets corresponds to initial conditions leading to successful
initiation, and the other to decay. In particular, if the initial
condition satisfies $u(x,0)<u\crit(x)$, $x\in[0,\infty)$ then $u(x,t)$ decays as
$t\to\infty$, and if $u(x,0)>u\crit(x)$, $x\in[0,\infty)$ then
$u(x,t)$ approaches a stable propagating front solution.
Moreover, if a continuous one-parametric
family of initial conditions contains some that initiate a wave and
some that lead to decay, then there is always at least one that does
neither, but gives a solution that approaches the critical nucleus.
This critical nucleus is the same for all such families, e.g. does not
depend on the shape of the initial distribution $u(x,0)$, as long as
its amplitude is at the threshold corresponding to that shape. Initial conditions
very close to the threshold generate solutions which approach the
critical nucleus and then depart from it, either toward propagation
or toward decay. This transient stationary state can be seen in
\fig{fhn}(c,d) where the initial conditions are selected very close to
the threshold.

%%%%%%%%%%%%%%%%%%%
\begin{figure*}[ht]
\includegraphics{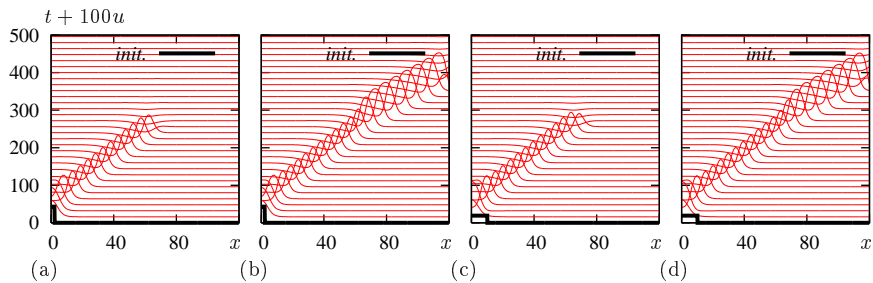}
\caption[]{
  The critical pulse is a universal transient for any near-threshold
  initial condition.
  The solutions to \eq(fhn1) for slightly below-threshold (a,c)
  and slightly above-threshold (b,d) amplitudes, for
  smaller stimulus width $\xstim=2.10$ in (a,b) and larger
  $\xstim=10.05$ in (c,d).
  Parameter values:
  $\beps=0.02$,
  $\alpha =0.37$,
  $\dt =0.01$,
  $\dx =0.15$,
  $L=120$.
  Stimulus amplitudes:
  $\ustim=0.431929399574766$ for (a),
  $0.431929399574768$ for (b),
  $0.191802079312694$ for (c) and
  $0.191802079312696$ for (d).
  In all cases we see a slow, low-amplitude unstable propagating pulse
  which subsequently either decays or evolves into a fast,
  high-amplitude stable propagating pulse.
}\label{fig:critpulse}
\end{figure*}
%%%%%%%%%%%%%%%%%%%

For small $\beps>0$, system \eq(fhn1) does not have nontrivial
stationary solutions.
However, for $x\in(-\infty,\infty)$ equation \eq(fhn1)
has an unstable
propagating pulse solution $\tilde{u}\crit(x-ct)$,
$\tilde{v}\crit(x-ct)$ such that $\tilde{u}\crit(x)\to u\crit(x)$,
$\tilde{v}\crit(x)\to0$ and $c=\O{\beps^{1/2}}$ as
$\beps\searrow0$.
Due to translational symmetry, this solution (in the comoving frame of reference)
has a zero eigenvalue
corresponding to the eigenfunction $(\tilde{u}\crit',\tilde{v}\crit')$.
This solution also has a single unstable eigenvalue.
So its center-stable
manifold has codimension one and is the threshold hypersurface
dividing the phase space into the decay domain and the initiation
domain. So here we have a \emph{critical pulse} solution, which we define as
an unstable traveling wave that is asymptotic to the resting
state for both limits $x-ct\to\pm\infty$.
For small $\beps$, the critical pulse is
essentially a slowly traveling variant of the critical nucleus. Any
solution with the initial condition at the threshold hypersurface will
asymptotically approach this critical pulse (suitably shifted),
and any solutions starting close to the threshold will approach this
critical pulse as a transient~\endnote{
  The symmetry $x \leftrightarrow -x$ means there are two stable pulse solutions,
  one propagating to the right and one propagating to the left.
  Likewise, there are two critical pulse solutions, and two
  center-stable manifolds.  The relationship between these two
  critical hypersurfaces is complicated, since some families of
  initial conditions can generate two oppositely traveling pulses
  and some can generate only one.
}. This is illustrated in \fig{critpulse}.

With this understanding, the excitation condition in terms of
$(\xstim,\ustim)$ reduces to computing the intersection of the
two-parametric manifold described by \eq(fhn1i) with the codimension 1
stable (center-stable) manifold of the critical nucleus (critical
pulse). This gives the curve on the $(\xstim,\ustim)$ plane separating
initial conditions leading to excitation propagation from those
leading to decay.  This can be done numerically or,
with appropriate simplifications, analytically. An example of dealing
with this problem in the ZFK equation, using Galerkin style
approximations can be found in \cite{Neu-etal-1997}.
Here we concentrate on the principal question, about the nature of the
threshold hypersurface in the functional space, for cardiac excitation
equations. It appears that in cardiac equations, this nature is
different from the FitzHugh-Nagumo theory just considered.

%##############################
\section{Simplified cardiac excitation model}

Now we consider the simplified model of $\INa$-driven excitation
fronts in typical cardiac excitation models proposed in
\cite{Biktashev-2002}:
\begin{equation}
E_t = E_{xx}+\Heav(E-1)h, \quad
h_t = (\Heav(-E)-h)/\tau	\label{eq:bik}
\end{equation}
with boundary condition
\begin{equation}
E_x(0,t) =0 \label{eq:bik1b}
\end{equation}
and initial condition
\begin{eqnarray}
E(x,0) = -\alpha+\Estim\Heav(\xstim-x), \quad
h(x,0) = 1, \label{eq:bik1a}
\end{eqnarray}
where the variable $E$ represents transmembrane potential of the
cardiac tissue, $h$ is the probability of the Na-gates being
open, $\tau$ is a dimensionless parameter and $\alpha>0$ represents
the pre-frontal voltage which we consider fixed in this paper.  System
\eq(bik) can be obtained by simplifying right-hand sides of the
\emph{fast} subsystem in an appropriate asymptotic limit of a typical
cardiac excitation model \cite{Biktasheva-etal-2006}.  In that sense,
system \eq(bik) plays the same role for a typical cardiac excitation model,
as ZFK equation \eq(zfk) plays for a classical activator-inhibitor
excitable system like \eq(fhn1).

System \eq(bik) does not have nontrivial
bounded stationary solutions: if $E_t=h_t=0$ then any bounded solution
has the form $E=a$, $h=\Heav(-a)$, $a=\const$. So, \emph{there are no
critical nuclei} in this system.
  Nevertheless, system \eq(bik) gives propagating front solutions for
  initial conditions above a threshold and decay for those below it.
Hence there is a question, what happens when the
initial condition is exactly at the threshold.

System \eq(bik) has a family of propagating front solutions
\begin{eqnarray}
E(z) &=& \left\{\begin{array}{ll}
\displaystyle  \omega - \frac{\tau^2 c^2}{1+\tau c^2}\exp\left(\frac{z}{\tau c}\right) & (z\leq-\xm) , \\[1.5ex]
\displaystyle -\alpha + \alpha \exp(-cz)	& (z\geq-\xm),
\end{array}\right. 				\nonumber \\ \nonumber\\
h(z) &=& \left\{\begin{array}{ll}
\displaystyle \exp\left( \frac{z}{\tau c} \right)	& (z\leq 0) , \\
\displaystyle 1	& (z\geq 0) ,
\end{array}\right.				\label{eq:Ehprofile}
\end{eqnarray}
where $z=x-ct$,
$\omega=1+\tau c^2(\alpha+1)$,
$\xm=\frac{1}{c}\ln\left(\frac{1+\alpha}{\alpha}\right)$ and
parameters $c$, $\alpha$ and $\tau$ related by
\begin{equation}
\tau c^2 \ln\left((1+\alpha)(c^2+\tau^{-1})\right) +
\ln\left(1+\alpha^{-1}\right)=0.
							\label{eq:transalpha}
\end{equation}
For a fixed $\alpha$, there is a $\tau_*(\alpha)$ such that for
$\tau>\tau_*$, equation \eq(transalpha) has two solutions for $c$,
$c=c_{\pm}(\alpha,\tau)$, $c_+>c_-$
\cite{Biktashev-2002}. There is numerical and analytical evidence that solutions
with $c=c_+$ are stable and those with $c=c_-$ are unstable with one
positive eigenvalue \cite{Biktashev-2002,Hinch-2004}.

Hence by analogy with the FHN system, we propose the following
\begin{conjecture}
The center-stable manifold of the unstable front solution
\eq(Ehprofile) with $c=c_-(\alpha,\tau)$
is the threshold hypersurface, separating the initial conditions leading to
initiation from the initial conditions leading to decay.\footnote{
  Again, symmetry $x\leftrightarrow-x$
  implies there are actually two hypersurfaces, partly connected with
  each other.
}
\end{conjecture}

That is, instead of a critical nucleus or a critical pulse solution,
the role of the threshold solution is played by a ``critical front'',
which we define as a traveling wave solution with different asymptotics
at $x-ct\to+\infty$ and $x-ct\to-\infty$: the pre-frontal state and the post-frontal state.

An ``experimentally testable'' consequence of this conjecture is that
for any initial condition exactly at the threshold, the solution will
approach the unstable front as $t\to+\infty$. For any initial
condition near the threshold, the solution will come close to the
unstable front and stay in its vicinity for a long time: if the
positive eigenvalue is $\lambda$ and the initial condition is
$\delta$-close to the threshold, the transient front should be
observed for the time of the order of $\lambda^{-1}|\ln\delta|$.  This
transient front solution \emph{will not depend on the initial
condition}, as long as the initial condition is at the threshold.

%%%%%%%%%%%%%%%%
\begin{figure}[ht]
\includegraphics{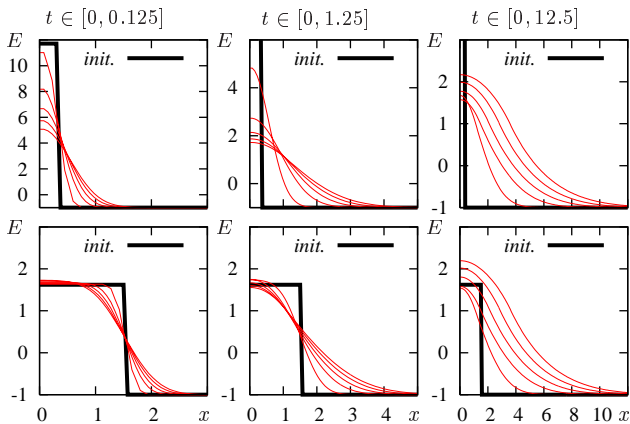}
\caption[]{
  Evolution of two different near-threshold initial conditions toward the critical
  front solution in system \eq(bik).
  Initial stimuli:
  $\xstim=0.3$, $\Estim=12.716330706144868$ (upper row)
  and
  $\xstim=1.5$, $\Estim=2.619968799545055$ (lower row).
  Other parameters:
  $\tau=8.2$, $\alpha=1$, $\dx=0.075$, $\dt=0.0025$, $L=50$.
}\label{fig:transient}
\end{figure}
%%%%%%%%%%%%%%%%

\begin{figure}[ht]
\includegraphics{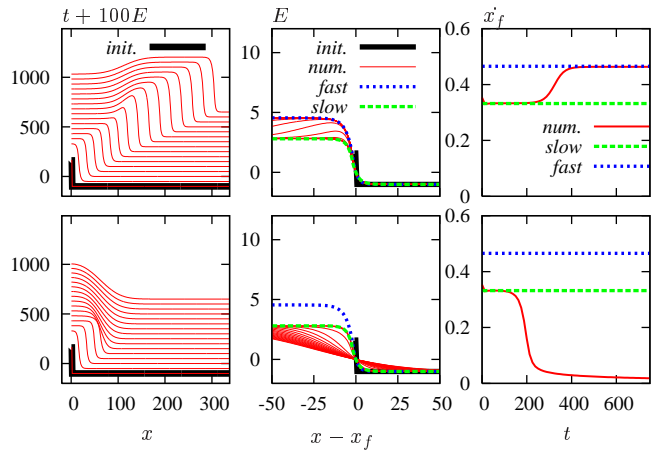}
\caption[]{
Transient critical fronts are close to the unstable front solution
of \eq(bik). Initial conditions: $\xstim=1.5$, with
$\Estim=2.619968799545055$ in the upper row and
$\Estim=2.619968799545054$ in the lower row,
other parameters the same as in \fig{transient}.  Left column: evolution
of the $E$ profiles in the laboratory frame of reference. Middle
column: same evolution, in the frame of reference comoving with the
front. Right column: speed of the front.  Blue/green dashed lines in the
middle and right columns correspond to the exact fast/slow front
solutions of \eq(bik).
}\label{fig:critfront}
\end{figure}

We have tested these predictions by numerical simulation of
\eqthree(bik,bik1b,bik1a). The results are shown in \figs{transient}
and~\ref{fig:critfront}.

\Fig{transient} illustrates two solutions starting from initial conditions
with different $\xstim$ values. In both cases, $\Estim$ values have been
chosen close to the respective thresholds with high precision. In both
cases, the solutions evolve in the long run toward the same
propagating front.

\Fig{critfront} presents an analysis of a pair of solutions, one with slightly
above-threshold and the other with slightly below-threshold initial
conditions. To separate the evolution of the front shape from its
movement, we employed the idea of symmetry group decomposition with
explicit representation of the orbit manifold (see
e.g. \cite{Biktashev-Holden-1998}). Practically, we
define the front point $\xf=\xf(t)$ via
\[
		E(\xf(t),t)=E_*
\]
for some constant $E_*$ which is guaranteed to be represented exactly
once in the front at every instant of time (we have chosen $E_*=0$).
Then $E(x-\xf(t),t)$ gives the voltage profile ``in the standard position'', and
$\xf(t)$ describes the movement of this profile.

%%%%%%%%%%%%%%%%
%%%%%%%%%%%%%%%%

The predictions based on the Conjecture are that the voltage profile
should, after an initial transient depending on the initial condition,
approach the profile of the slow unstable front solution given by
\eq(Ehprofile) with $c=c_-(\tau,\alpha)$ and stay close to it for some time, before
either developing into the fast stable front
\eq(Ehprofile) with $c=c_+(\tau,\alpha)$
or decaying. Likewise, the speed of the front should, after an initial
transient, be close to the speed of the slow unstable front
$c_-(\tau,\alpha)$, before either switching the speed of the fast
stable front $c_+(\tau,\alpha)$ or dropping to zero. This is precisely
what is seen on \fig{critfront}, where we have taken advantage of
knowing the exact solutions $E(x-c_{\pm}t)$ and $c_\pm$ for both the fast and
the slow fronts.

Initial conditions with different $\xstim$ and $\Estim$ close to the
corresponding threshold produce the same picture with the exception
of the initial transient. We have also checked that the length of the time
period during which the solution stays close to the unstable front is,
roughly, a linear function of the number of correct decimal figures in
$\Estim$, as it should be according to the Conjecture.

%##############################
\section{Detailed cardiac excitation model}

%%%%%%%%%%%%%%%%
\begin{figure}[ht]
\includegraphics{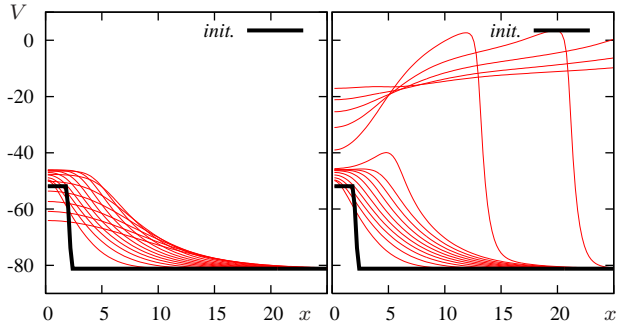}
\caption[]{
  Critical fronts in CRN model~\cite{Courtemanche-etal-1998}.
  Shown are voltage profiles in every $10\,\ms$.
  Parameter values:
  $\dt =0.01\,\ms$,
  $\dx =0.2$,
  $L=40$, the length unit chosen so that voltage diffusion coefficient equals 1.
  Stimulus witdh $\xstim=2$, stimulus amplitudes:
  $\Vstim=29.31542299307152\,\mV$ (left panel) and
  $\Vstim=29.31542299307153\,\mV$ (right panel).
  The critical fronts are formed within first $10\,\ms$ and then
  are seen for subsequent $80\,\ms$ on both panels before exploding into an excitation
  wave of much bigger amplitude and speed on the right panel, and decaying on the left panel.
}\label{fig:CRN}
\end{figure}
%%%%%%%%%%%%%%%%

The simplified model \eq(bik) is quantitatively very far from any
realistic ionic model of cardiac excitation, and has many peculiar
qualitative features, stemming from the non-standard asymptotic embedding
leading to it. Hence the newly described phenomenon of critical front
could be an artifact of the simplifications. To eliminate this
possibility, we have tested the relevance of the critical front
concept to a full ionic model of cardiac excitation. We have chosen
the model of human atrial tissue due to Courtemanche\ et\
al. (CRN)~\cite{Courtemanche-etal-1998}, which is less stiff than a
typical ventricular or Purkinje fiber model, is well formulated in a
mathematical sense and is popular among cardiac modellers.  This
model operates with 21 dynamic variables including the transmembrane
voltage $V$. We have used the default parameter values as described in
\cite{Courtemanche-etal-1998} and supplemented the equation for $V$ in the ODE system with a
diffusion term $D\partial^2V/\partial{x}^2$. As the spatial scale is
not important for the question at hand, we assumed $D=1$. The
initial conditions for $V$ were taken in the form
\[
  V(x,0) = V_r+\Vstim\Heav(\xstim-x)
\]
where $V_r=-81.18\,\mV$ is the standard resting potential, and for all
other 20 variables at their resting values as described in
\cite{Courtemanche-etal-1998}. \Fig{CRN} illustrates a pair of
solutions with initial conditions slightly above and slightly below
the threshold. The critical front solution is clearly seen there: it
has the upper voltage of about $-46\,\mV$ and during $80\,\ms$ of its
existence propagates with a speed approximately $0.06$ space units per
millisecond. Then for the above-critical case it develops into an
excitation front with maximal voltage about $+3\,\mV$ and speed 0.8
space units per millisecond, and decays for the below-critical case.

Mathematically, the post-front voltage of about $-46\,\mV$ observed in
\fig{CRN} is not a true equilibrium of the full CRN model so the
critical front can only be an asymptotic concept in an appropriate
asymptotic embedding, say as ones described in
\cite{Biktasheva-etal-2006} or
\cite{Simitev-Biktashev-2006}, and the observed critical front may well be
the front of a critical pulse solution in the full model.  However
\fig{CRN} demonstrates that the critical front is a practical and well
working concept even for the full model, unlike the critical pulse
which may be theoretically existing but practically unobservable:
notice the number of significant decimal digits in initial conditions
required to produce only the critical front observed for $80\,\ms$, and
recall that the number of decimals is roughly proportional to the
duration of the observation of an unstable solution.

%##############################
\section{Discussion}
We have presented numerical evidence that the
center-stable manifold of the unstable slow front solution of
the fast subsystem of a cardiac excitation model
serves as the threshold hypersurface separating initial conditions
leading to successful initiation and those leading to decay. This
means e.g. that a critical curve for a two-parametric family of
initial conditions, be that family \eq(bik1a) with parameters
$(\xstim,\Estim)$ or any other, can be found as an intersection of
this codimension-1 critical hyper-surface with the two-dimensional
manifold of those initial conditions. Finding this hypersurface can be
done numerically or analytically using suitable approximation, e.g. as
it was done in \cite{Neu-etal-1997,Moll-Rosencrans-1990} for the ZFK equation;
this is a subject for further investigation.

Another problem for future study is to verify that the findings
remain qualitatively true for formal asymptotic embeddings
of various cardiac excitation models, and in particular,
which mathematical features of these embeddings are essential for the existence
of the critical fronts.
Of principal
importance is the conclusion that for cardiac equations, instead of the
``critical nucleus'' or its slowly moving variant ``critical pulse'' known from
the FitzHugh-Nagumo theory, we now have a ``critical front solution''.
This, in particular, means physically that the make-or-break conditions of cardiac
excitation wave are restricted to a vicinity of its front.

Finally, we believe that this study sets a useful example for
initiation problems in other types of excitable systems, alternative
to the existing critical nucleus theory, since not all, if any,
real-world excitable systems are well described by an asymptotic
structure as in \eq(fhn1).

Authors are grateful to C.F.~Starmer and E.E.~Shnol for inspiring
discussion.  The study has been supported by EPSRC grant GR/S75314/01
and MacArthur Foundation grant 71356-01.

% \bibliography{cf} 

\begin{thebibliography}{19}
\expandafter\ifx\csname natexlab\endcsname\relax\def\natexlab#1{#1}\fi
\expandafter\ifx\csname bibnamefont\endcsname\relax
  \def\bibnamefont#1{#1}\fi
\expandafter\ifx\csname bibfnamefont\endcsname\relax
  \def\bibfnamefont#1{#1}\fi
\expandafter\ifx\csname citenamefont\endcsname\relax
  \def\citenamefont#1{#1}\fi
\expandafter\ifx\csname url\endcsname\relax
  \def\url#1{\texttt{#1}}\fi
\expandafter\ifx\csname urlprefix\endcsname\relax\def\urlprefix{URL }\fi
\providecommand{\bibinfo}[2]{#2}
\providecommand{\eprint}[2][]{\url{#2}}

\bibitem[{\citenamefont{Krinsky and Swinney}(1991)}]{Krinsky-Swinney-1991}
\bibinfo{editor}{\bibfnamefont{V.}~\bibnamefont{Krinsky}} \bibnamefont{and}
  \bibinfo{editor}{\bibfnamefont{H.}~\bibnamefont{Swinney}}, eds.,
  \emph{\bibinfo{title}{Wave and patterns in biological and chemical excitable
  media}} (\bibinfo{publisher}{North-Holland}, \bibinfo{address}{Amsterdam},
  \bibinfo{year}{1991}).

\bibitem[{\citenamefont{Zipes and Jalife}(2000)}]{Zipes-Jalife-2000}
\bibinfo{editor}{\bibfnamefont{D.~P.} \bibnamefont{Zipes}} \bibnamefont{and}
  \bibinfo{editor}{\bibfnamefont{J.}~\bibnamefont{Jalife}}, eds.,
  \emph{\bibinfo{title}{Cardiac electrophysiology: From cell to bedside}}
  (\bibinfo{publisher}{W B Saunders Co}, \bibinfo{year}{2000}).

\bibitem[{\citenamefont{FitzHugh}(1961)}]{FitzHugh-1961}
\bibinfo{author}{\bibfnamefont{R.}~\bibnamefont{FitzHugh}},
  \bibinfo{journal}{Biophysical Journal} \textbf{\bibinfo{volume}{1}},
  \bibinfo{pages}{445} (\bibinfo{year}{1961}).

\bibitem[{\citenamefont{Nagumo et~al.}(1962)\citenamefont{Nagumo, Arimoto, and
  Yoshizawa}}]{Nagumo-etal-1962}
\bibinfo{author}{\bibfnamefont{J.}~\bibnamefont{Nagumo}},
  \bibinfo{author}{\bibfnamefont{S.}~\bibnamefont{Arimoto}}, \bibnamefont{and}
  \bibinfo{author}{\bibfnamefont{S.}~\bibnamefont{Yoshizawa}},
  \bibinfo{journal}{Proc. IRE} \textbf{\bibinfo{volume}{50}},
  \bibinfo{pages}{2061} (\bibinfo{year}{1962}).

\bibitem[{\citenamefont{Rushton}(1937)}]{Rushton-1937}
\bibinfo{author}{\bibfnamefont{W.}~\bibnamefont{Rushton}},
  \bibinfo{journal}{Proc. Roy. Soc. Lond. ser. B}
  \textbf{\bibinfo{volume}{124}}, \bibinfo{pages}{210} (\bibinfo{year}{1937}).

\bibitem[{\citenamefont{Starmer et~al.}(2003)\citenamefont{Starmer, Grant, and
  Colatsky}}]{Starmer-etal-2003}
\bibinfo{author}{\bibfnamefont{C.~F.} \bibnamefont{Starmer}},
  \bibinfo{author}{\bibfnamefont{A.~O.} \bibnamefont{Grant}}, \bibnamefont{and}
  \bibinfo{author}{\bibfnamefont{T.~J.} \bibnamefont{Colatsky}},
  \bibinfo{journal}{Cardiovasc. Res.} \textbf{\bibinfo{volume}{57}},
  \bibinfo{pages}{1062} (\bibinfo{year}{2003}).

\bibitem[{\citenamefont{Biktashev and Suckley}(2004)}]{Biktashev-Suckley-2004}
\bibinfo{author}{\bibfnamefont{V.~N.} \bibnamefont{Biktashev}}
  \bibnamefont{and} \bibinfo{author}{\bibfnamefont{R.}~\bibnamefont{Suckley}},
  \bibinfo{journal}{Phys. Rev. Lett.} \textbf{\bibinfo{volume}{93}},
  \bibinfo{pages}{168103} (\bibinfo{year}{2004}).

\bibitem[{\citenamefont{Flores}(1989)}]{Flores-1989}
\bibinfo{author}{\bibfnamefont{G.}~\bibnamefont{Flores}}, \bibinfo{journal}{J.
  Diff. Eq.} \textbf{\bibinfo{volume}{80}}, \bibinfo{pages}{306}
  (\bibinfo{year}{1989}).

\bibitem[{\citenamefont{Flores}(1991)}]{Flores-1991}
\bibinfo{author}{\bibfnamefont{G.}~\bibnamefont{Flores}},
  \bibinfo{journal}{SIAM J. Math. Anal.} \textbf{\bibinfo{volume}{22}},
  \bibinfo{pages}{392} (\bibinfo{year}{1991}).

\bibitem[{\citenamefont{Zel'dovich and
  Frank-Kamenetsky}(1938)}]{Zeldovich-FrankKamenetsky-1938}
\bibinfo{author}{\bibfnamefont{Y.~B.} \bibnamefont{Zel'dovich}}
  \bibnamefont{and} \bibinfo{author}{\bibfnamefont{D.~A.}
  \bibnamefont{Frank-Kamenetsky}}, \bibinfo{journal}{Doklady AN SSSR}
  \textbf{\bibinfo{volume}{19}}, \bibinfo{pages}{693} (\bibinfo{year}{1938}).

\bibitem[{\citenamefont{McKean}(1970)}]{McKean-1970}
\bibinfo{author}{\bibfnamefont{H.~P.} \bibnamefont{McKean},
  \bibfnamefont{Jr.}}, \bibinfo{journal}{Adv. Appl. Math.}
  \textbf{\bibinfo{volume}{4}}, \bibinfo{pages}{209} (\bibinfo{year}{1970}).

\bibitem[{\citenamefont{Neu et~al.}(1997)\citenamefont{Neu, Preissig, and
  Krassowska}}]{Neu-etal-1997}
\bibinfo{author}{\bibfnamefont{J.~C.} \bibnamefont{Neu}},
  \bibinfo{author}{\bibfnamefont{R.~S.} \bibnamefont{Preissig}},
  \bibnamefont{and}
  \bibinfo{author}{\bibfnamefont{W.}~\bibnamefont{Krassowska}},
  \bibinfo{journal}{Physica D} \textbf{\bibinfo{volume}{102}},
  \bibinfo{pages}{285} (\bibinfo{year}{1997}).

\bibitem[{\citenamefont{Biktashev}(2002)}]{Biktashev-2002}
\bibinfo{author}{\bibfnamefont{V.~N.} \bibnamefont{Biktashev}},
  \bibinfo{journal}{Phys. Rev. Lett.} \textbf{\bibinfo{volume}{89}},
  \bibinfo{pages}{168102} (\bibinfo{year}{2002}).

\bibitem[{\citenamefont{Biktasheva et~al.}(2006)\citenamefont{Biktasheva,
  Simitev, Suckley, and Biktashev}}]{Biktasheva-etal-2006}
\bibinfo{author}{\bibfnamefont{I.~V.} \bibnamefont{Biktasheva}},
  \bibinfo{author}{\bibfnamefont{R.~D.} \bibnamefont{Simitev}},
  \bibinfo{author}{\bibfnamefont{R.~S.} \bibnamefont{Suckley}},
  \bibnamefont{and} \bibinfo{author}{\bibfnamefont{V.~N.}
  \bibnamefont{Biktashev}}, \bibinfo{journal}{Phil. Trans. Roy. Soc. Lond. ser.
  A} \textbf{\bibinfo{volume}{364}}, \bibinfo{pages}{1283}
  (\bibinfo{year}{2006}).

\bibitem[{\citenamefont{Hinch}(2004)}]{Hinch-2004}
\bibinfo{author}{\bibfnamefont{R.}~\bibnamefont{Hinch}},
  \bibinfo{journal}{Bull. Math. Biol.} \textbf{\bibinfo{volume}{66}},
  \bibinfo{pages}{1887} (\bibinfo{year}{2004}).

\bibitem[{\citenamefont{Biktashev and Holden}(1998)}]{Biktashev-Holden-1998}
\bibinfo{author}{\bibfnamefont{V.~N.} \bibnamefont{Biktashev}}
  \bibnamefont{and} \bibinfo{author}{\bibfnamefont{A.~V.}
  \bibnamefont{Holden}}, \bibinfo{journal}{Physica D}
  \textbf{\bibinfo{volume}{116}}, \bibinfo{pages}{342} (\bibinfo{year}{1998}).

\bibitem[{\citenamefont{Courtemanche et~al.}(1998)\citenamefont{Courtemanche,
  Ramirez, and Nattel}}]{Courtemanche-etal-1998}
\bibinfo{author}{\bibfnamefont{M.}~\bibnamefont{Courtemanche}},
  \bibinfo{author}{\bibfnamefont{R.~J.} \bibnamefont{Ramirez}},
  \bibnamefont{and} \bibinfo{author}{\bibfnamefont{S.}~\bibnamefont{Nattel}},
  \bibinfo{journal}{Am. J. Physiol.} \textbf{\bibinfo{volume}{275}},
  \bibinfo{pages}{H301} (\bibinfo{year}{1998}).

\bibitem[{\citenamefont{Simitev and Biktashev}(2006)}]{Simitev-Biktashev-2006}
\bibinfo{author}{\bibfnamefont{R.~D.} \bibnamefont{Simitev}} \bibnamefont{and}
  \bibinfo{author}{\bibfnamefont{V.~N.} \bibnamefont{Biktashev}},
  \bibinfo{journal}{Biophysical Journal} \textbf{\bibinfo{volume}{90}},
  \bibinfo{pages}{2258} (\bibinfo{year}{2006}).

\bibitem[{\citenamefont{Moll and Rosencrans}(1990)}]{Moll-Rosencrans-1990}
\bibinfo{author}{\bibfnamefont{V.}~\bibnamefont{Moll}} \bibnamefont{and}
  \bibinfo{author}{\bibfnamefont{S.~I.} \bibnamefont{Rosencrans}},
  \bibinfo{journal}{SIAM J. Appl. Math.} \textbf{\bibinfo{volume}{50}},
  \bibinfo{pages}{1419} (\bibinfo{year}{1990}).

\end{thebibliography}

\end{document}